\documentstyle[prl,aps,epsf]{revtex}

\begin{document}
\newcommand{\be}{\begin{equation}}
\newcommand{\ee}{\end{equation}}
\newcommand{\bq}{\begin{eqnarray}}
\newcommand{\eq}{\end{eqnarray}}

\title{On the Structure of the Bose-Einstein Condensate Ground State} 
\author{
A. I. Solomon\footnote{Email: a.i.solomon@open.ac.uk},  $\;$ Y. Feng\footnote{Email: y.feng@open.ac.uk}, $\;$ and $\;$   V. Penna\footnote{Permanent address: Dipartimento di Fisica, Politecnico di Torino, Corso Duca degli Abruzzi 24, I-10129 Torino, Italy. Email: penna@athena.polito.it}\\
{\small \sl  Faculty of Mathematics and Computing,}\\
{\small \sl  The Open University, Milton Keynes, MK7 6AA, United Kingdom } \\
}
\date{\today}
\maketitle

\begin{abstract}
We construct a  macroscopic wave function 
that describes the Bose-Einstein condensate and weakly excited states, using the $su(1,1)$ structure of the mean-field hamiltonian, and compare this state with the experimental values of second and third order correlation functions.
\end{abstract}
%
%

The recent experimental achievement\cite{Anderson} of Bose-Einstein condensation (BEC) has stimulated a great revival of interest in the theoretical study of this phenomenon. One fascinating aspect of the Bose-Einstein Condensate is the nature of coherence in a macroscopic quantum system, and in recent experiments some of the coherence properties of BEC have been discussed and explicitly addressed\cite{Kett1,Kett2,Burt}.  In this paper, we describe the hamiltonian and energy eigenstates within the su(1,1) mean-field picture of BEC  and, based on this theory, we construct a generalised version of the BEC weakly excited states. We calculate some correlation functions within this theory, and compare with recent experimental results. 

The standard description of the Bose-Einstein condensate is by means of
an order parameter field $\Psi ({\bf x})$ which  accounts locally
for the physical state of the system\cite{BE}. The hamiltonian has the
standard form
\bq
{\cal H} [\Psi] \;& =& \int d^3 x  
\left [
{\frac{\hbar^2}{2m}} |\nabla\Psi({\bf x}) |^2
+ U({\bf x}) |\Psi ({\bf x})|^2  \right ] \nonumber \\
&&{}+ {\frac{1}{2}}
\int d^3 x \int d^3 y \, \Psi^* ({\bf y}) \,\Psi^*({\bf x}) 
V({\bf x} , {\bf y}) \,\Psi ({\bf y}) \,\Psi({\bf x}) \, . \label{GPE}
\eq
Representing the field $\Psi ({\bf x})$ by its
Fourier transformation leads to the second quantized form 
%
\be
{\cal H} = \sum_k\epsilon_k  n_k +
{\frac{1}{2}} \sum_k \sum_{p,q}\, 
V_k a^{+}_{p+k} a^{+}_{q-k} \, a_p \, a_q \; ,
\label{H1}
\ee
where $V_k \equiv V(|{\bf k}|)$, is a momentum preserving interaction. The number operators $n_k\equiv a^{+}_k a_k$,
the raising operators $a^{+}_k$, and the lowering operators
$a_{l}$ obey the Weyl-Heisenberg algebra commutators
\be
[a_l, a^{+}_k] =  \delta_{lk} {\bf I} \, ,\;    
[a^{+}_l , n_k ] = - \delta_{lk} a^{+}_k   \, ,\;  
[a_l, n_k ] = \delta_{lk} a_k  \;.
\label{A1}
\ee

The Bogoliubov prescription is that at zero temperature the state with
$k=0$ is macroscopically occupied and this observation allows one to treat
$a^{+}_0$ and $a_0$ as c-numbers ($[a_0, a^{+}_0] \simeq 0$)
since the corresponding number operator $ n_0$,
counting the bosons constituting the condensate,
turns out to be macroscopically large. However this neglect of the operators $a^{+}_0$ and $a_0$ is not an appropriate approximation if we wish to describe phenomena in the condensate ground states. So here we no longer adopt such an approximation and we retain the  operator status of $a^{+}_0$ and $a_0$ in order to give a more consistent description of the state of the condensed system.

Making explicit the terms depending on $a_0$ and $a_0^+$

in (\ref{H1}) and neglecting those terms that contain three
or four boson operators $a^{+}_k$, $a_l$ ($k, l \ne 0$) reduces
${\cal H}$ to the form
\bq
{\cal H} &= &\epsilon_0 n_0 + {\frac{V_0}{2}} a^{+2}_0a^{2}_0 +
\sum_{k\ne 0} \epsilon_k  n_k + n_0 V_0 \sum_{k\ne 0} \, n_k \nonumber\\
&&{}+{\frac{1}{2}} \sum_{k\ne 0}
 V_k[n_0 \,( n_k+n_{-k}) + a^2_{0} \, a^+_{k} a^+_{-k} + 
a^{+2}_{0}\,a_{k} a_{-k}] 
\eq
where $\epsilon_k = \hbar^2 k^2/2m$ ($m$ is the fluid atom mass).

The  hamiltonian is linearized by using the mean-field approximation procedure which reduces bilinear operators such as $A B$ to the linear form
$$
AB \simeq A \langle B \rangle +  \langle A \rangle B
-\langle A \rangle \langle B \rangle
$$
based on the assumption
$(A - \langle A \rangle )(B -\langle B \rangle ) \simeq 0$.
We note that a similar approach, starting instead  from a bosonic Hubbard model and using a novel form of this linearisation procedure \cite{Luigi}, leads to essentially to the same $su(1,1)$ structure for the excited states, as well as an additional condensate term for the ground state. This yields the quadratic reduced hamiltonian
%
$$
{\cal H}_{mf} = \sigma_0( n_0+\frac{1}{2}) +{\frac{1}{2}}( u_0 a^{+2}_{0}  +
u^{*}_0\, a^2_{0} ) + \sum_{k\ne 0} {\cal H}^{(k)} -E_*
$$
where
$$
\sigma_0 \equiv \epsilon_0 +\frac{1}{2}\sum_{k\ne 0}  (V_0 + V_k) (\langle n_k \rangle+\langle n_{-k} \rangle) , 
\;\;\;
u_0 \equiv V_0 \langle a^2_{0} \rangle 
+ \sum_{k\ne 0} V_k \langle a_k a_{-k} \rangle 
$$
$$
E_* = \frac{1}{2}[ V_0  |\langle a^{+2}_{0}\rangle |^2+\sigma_0]
+\frac{1}{2}\sum_{k\ne 0} \,[ (\sigma_k -\epsilon_k) \langle n_k +  n_{-k}\rangle+\sigma_k] 
$$
$$
+{\frac{1}{2}} \sum_{k\ne 0} (
u_k \langle a^+_{k} a^+_{-k} \rangle +
u^{*}_k \langle a_{k} a_{-k}\rangle)\, ,
$$
and the pair mode hamiltonian ${\cal H}^{(k)}$ is
$$
{\cal H}^{(k)} ={\frac{\sigma_k}{2}} (n_k +n_{-k}+1) +
{\frac{1}{2}}( u_k a^{+}_{k} a^{+}_{- k} + u^{*}_k\, a_{k} a_{-k} )
$$
with
$$
\sigma_k \equiv \epsilon_k + \langle n_0 \rangle (V_0 + V_k),\,
u_k \equiv V_k \langle a^2_{0} \rangle .
$$
We can rewrite ${\cal H}_{mf}$ in the $su(1,1)$ form
\bq
{\cal H}_{mf} &=& 2\left [ \sigma_{0} A_3^{(0)} +
{\frac{1}{2}}  ( u_0 A_+^{(0)} + u^{*}_0\, A_-^{(0)}) \right ] \nonumber\\
&&{}+  \sum_{k\ne 0} \left [\sigma_{k} A_3^{(k)} +
{\frac{1}{2}}  ( u_k A_+^{(k)} + u^{*}_k\, A_-^{(k)}) \right ]-E_{*}
\label{SUH}
\eq
by means of the generators of the algebra $su(1,1)$ 
\be
A_3^{(0)}= \frac{1}{2}(n_0+ \frac{1}{2}),\,\, A_+^{(0)}= \frac{a^{+2}_0}{2},\,\, A_-^{(0)} =\frac{a^2_0}{2}\;
\ee
and
\be
A_3^{(k)} ={\frac{1}{2}} (n_{k} + n_{-k} +1), \,\, 
A_+^{(k)}= a^+_{k} a^+_{-k} ,\,\,
A_{-}^{(k)} = a_{k} a_{-k} \; 
\label{B1}
\ee
that account for the momentum creation/destruction processes
occurring in the fluid and involving the modes $k$ and $-k$.
These satisfy the usual  commutation relations
\be
[ A_+^{(q)} , A_-^{(q)} ] =- 2 A_3^{(q)}, \,
[ A_3^{(q)} , A_{\pm}^{(q)}] = {\pm} A_{\pm}^{(q)}, \,\, q=0,k,-k
\label{COM}
\ee

It is known that within the $su(1,1)$ mean-field picture
the energy eigenstates are expressed a direct product of
$su_k(1,1)$ coherent states\cite{ALLAN}. We therefore write the eigenstates as
\be
|\xi \rangle = |\xi_0 \rangle \otimes_{k \ne 0} | \xi_k \rangle
\ee
where
$$
|\xi_k \rangle = {\exp}[ \xi_k A_+^{(k)} -\xi_k^{*} A_-^{(k)}]
|0 \rangle 
$$
with ${\rm th}\, \xi_k = -u_k / \sigma_k$.  The
eigenvalues of ${\cal H}_k $ are given by $E_k=  \sqrt{ \sigma_k^2- |u_k|^2}$.

The factor $|\xi_0 \rangle $
is normally absent (i.e., it is implicitly traced away)
in the standard approach due to the semiclassical status
of $ a^+_0$,  $a_0$. Here it restores the condensate to
its role as a dynamically active degree of freedom, that is
$$
|\xi_0 \rangle = {\exp}[ \xi_0 A_+^{(0)} -\xi_0^* A_-^{(0)}]
|0 \rangle 
$$
where $ {\rm th}\, \xi_0= -u_0 / \sigma_0$. \\
Writing for brevity
\be
S(\xi_k ) \equiv {\exp}[ \xi_k A_+^{(k)} -\xi_k^{*} A_-^{(k)}]\nonumber
\ee
we may express the state $ | \xi \rangle $ as 
\be
| \xi \rangle =\otimes_{q} S( \xi_q)|0 \rangle \;\;(q=0,+k,-k)\nonumber
\ee
The operators $ S(\xi_k) $ are similar to, but not identical with, the vacuum squeezing operators $  \exp [\frac{1}{2}(\xi a^{+2}-\xi^{*} a^2)] $ familiar from Quantum Optics.

The structure of the state $|\xi \rangle$ clearly exhibits the imprint of
the mean-field dynamical algebra ${\cal A}_* =\oplus_k su_k(1,1)$ which
provides an approximate description of the dynamical processes occurring
inside the system. The main unattractive feature is the fact that
$$
\langle a_0 \rangle \equiv \langle \xi | a_0 | \xi \rangle =0
$$
(arising from the two-boson character of ${\cal A}_*$), whereas
the low temperature regime should be characterised by a nonvanishing
 order parameter
$\langle \Psi({\bf x}) \rangle $, or equivalently $ \langle a_0 \rangle \sqrt{V}$
(recall that $|\langle a_0 \rangle/\sqrt{V}|^2 \simeq N$ where $N$ is the total
particle number inside the volume $V$) due to the strong
depletion of the $k$ mode states. In the state $ | \xi \rangle $ we also clearly have $\langle a_k \rangle \equiv \langle \xi | a_k | \xi \rangle =0$.

In addition, the values of the  second order correlation function 
$$
g^{(2)}(0)=\frac{\langle a^{+2}_0a^2_0 \rangle}{{\langle a^+_0a_0 \rangle}^2}
$$
and the third order correlation function 
$$
g^{(3)}(0)=\frac{\langle a^{+3}_0a^3_0 \rangle}{{\langle a^+_0a_0 \rangle}^3}
$$  
for the states $|\xi \rangle$  do not agree with the experimental results, which seem to indicate  that $g^{(2)}(0)$ and $g^{(3)}(0) $ are not exactly equal to 1, but slightly larger than one\cite{Kett2,Burt}. However it is easy to  show that  $g^{(2)}(0)= 1$ and $g^{(3)}(0) = 1$ in the state $ D(\alpha)|0 \rangle  $($D$ state) if the mean density $ \langle n_0 \rangle $ is a large number, where $ D(\alpha) =  \exp (\alpha a_0^{+}-\alpha^{*} a_0) $.  

These considerations motivate our attempt to generalise $|\xi \rangle$
 to $ |\xi,z \rangle $
\be
|\xi,z \rangle\ = |\xi_0,z_0 \rangle \otimes_{k \ne 0} | \xi_k ,z_k\rangle
\ee
by introducing the further definitions
\bq
|\xi_0,z_0 \rangle &=& D(z_0) | \xi_o \rangle \nonumber\\
|\xi_k, z_k \rangle& = &D(z_k) | \xi_k \rangle \; ,
\eq
where $D(z_q)= {\exp}(z_q a^+_q -z^*_q a_q),\;\; q=0,k,-k$.

We now describe the BEC states by $ | \xi, z \rangle $  where 
\be
|\xi, z \rangle =\otimes_{q} |\xi_q, z_q \rangle = \otimes_{q}D(z_q)S(\xi_q)|0 \rangle \;\; (q=0,\pm 1, \pm 2 ...)\label{DS}
\ee
For obvious reasons, we refer to the state $ |\xi, z \rangle $ as a $DS$ state, the $DS$ operator being similar to, but not identical with, that  which produces a squeezed state in Quantum Optics, namely
$$
 {\exp} (z a^{+} -z^{*} a)  {\exp} [\frac{1}{2}(\xi a^{+2}-\xi ^{*} a^2)]
$$
(for a single mode).

The BEC state (\ref{DS}) involves a large number of parameters  \{$\xi_k, z_k $\} which, as is usual in mean-field theories, may in principle be determined by a self-consistent treatment. However, we would expect $ z_k =0 $ for $ k \ne 0 $ (since there is no condensation other than in the $ k=0$ state); and if we are primarily interested in condensate properties we need determine only $ \xi_0$ and $ z_0 $ (4 real parameters). These may be calculated from the condensate conditions, as we now show. We have the following expectations:
\bq
\langle \xi,z | a_0 |\xi,z  \rangle
&=& \langle \xi_o | D^+(z_0) a_0 D(z_0) | \xi_o \rangle
= z_0 \nonumber\\
\langle \xi,z | n_0 | \xi,z \rangle
&=& \langle \xi_0 | D^+(z_0) n_0 D(z_0) | \xi_o \rangle
= |z_0|^2 +{\rm sh}^2| \xi_0|
\nonumber\\
\langle \xi,z | a_k | \xi,z \rangle
&=& \langle \xi_k| a_k |\xi_k \rangle  = z_k \\
\langle \xi,z | a_k a_{-k} | \xi,z \rangle
&=& \langle \xi_k| a_k a_{-k} |\xi_k \rangle
= z_k z_{-k} \nonumber\\
\langle \xi,z | n_k | \xi,z \rangle
&=& \langle \xi_k|n_k |\xi_k \rangle =
|z_k|^2 + {\rm sh}^2 |\xi_k| \nonumber
\eq

The state  $| \xi,z \rangle$ incorporates both the $su(1,1)$ structure inherited from the spectrum-generating algebra approach to the mean field hamiltonian, as well as the  nonvanishing expectation values for the operators $a_k$ implicit in a conventional (Heisenberg-Weyl) coherent state. As we shall show, a  choice of the parameters  for the state  $| \xi,z \rangle$ state allows one to fit the experimental values of  $g^{(2)}(0)$ and $g^{(3)}(0) $ .

Ketterle and Miesner\cite{Kett2} pointed out that data on the condensate expansion energy, combined with spectroscopic scattering length measurements, can be used to give the second order correlation function $g^{(2)}(0)$ in alkali condensates. An experiment on a BEC of sodium\cite{Mewes,Tiesi} by Ketterle et al yielded $g^{(2)}(0)=1.25 \pm 0.58$, and the experiment on a rubidium condensate\cite{Holland} yielded  $g^{(2)}(0)=1.0 \pm 0.2$. In another important experiment\cite{Burt}, Burt et al. recently compared the trap loss due to three-body recombination of a rubidium condensate to that of a thermal cloud, and obtained $ 7.4 \pm 2.6 $ for the ratio of the third order correlation function $g^{(3)}(0)$ values in the thermal and condensed states. 

Although the experimental results are not inconsistent with a pure $D$ state, at least in the case of rubidium, indications for sodium are that  $g^{(2)}(0)$ and $g^{(3)}(0) $ are larger than 1. From the structure of the states $ |\xi,z\rangle $, we can see that the BEC ground state is
$$
| \xi_0,z_0 \rangle =D(z_0)S(\xi_0)|0 \rangle
$$
where
\be
D(z_0)={\exp}(z_0 a^+_0-z^*_0 a_0)
\ee
\be
S(\xi_0)={\exp}(\xi_0 A_+^{(0)} -\xi_0^* A_-^{(0)}),\, \xi_0=r \; {\exp}(i\phi)
\ee
with $A_+^{(0)}= \frac{a^{+2}_0}{2}$, $A_-^{(0)} =\frac{a^2_0}{2}$.
We now show how to choose the parameters of  the  BEC ground state $ |\xi_0,z_0 \rangle $ to fit the experimental values of   $g^{(2)}(0)$ and $g^{(3)}(0) $  cited above\cite{Kett2,Burt}.

The unitary transformation of the operators $ a_0 $ and $ a^+_0 $ by $ D(z_0) $ and $ S(\xi_0)$ is given by
\bq
S^+(\xi_0)D^+(z_0)a_0D(z_0)S(\xi_0)=\mu a_0 + \nu a^+_0+z_0\nonumber\\
S^+(\xi_0)D^+(z_0)a^+_0D(z_0)S(\xi_0)=\mu a^+_0 + \nu^* a_0+z^*_0
\eq
where we have put
$$
\mu={\rm ch}r, \,\, \nu={\exp}(i\phi){\rm sh}r
$$
We  obtain the following mean values in the $DS$ state:
\bq
\langle  n_0 \rangle &=& |z_0|^2 +| \nu| ^2 \nonumber\\
\langle a^{+2}_0a^2_0 \rangle &=&3|\nu|^4 + 4|z_0|^2|\nu|^2 + |z_0|^4 + |\nu|^2 + \mu(z^2_0 \nu^* + z_0^{*2}\nu)\\
\langle a^{+3}_0a^3_0 \rangle &=& 15|\nu|^6 +27|\nu|^4|z_0|^2 + 9|\nu|^2|z_0|^4 + |z_0|^6+ 9|\nu|^4 \nonumber\\
&&{}+ 9|\nu|^2|z_0|^2 + 3(|z_0|^2+3|\nu|^2)[\mu(z^2_0 \nu^* + z_0^{*2}\nu)] \nonumber
\eq
If we write
$$
\mu(z^2_0 \nu^* + z^{*2}_0\nu)=|z_0|^2 \mu |\nu|\cos(\phi-\phi_z), \, z_0=|z_0|{\exp}(i\phi_z)
$$
Then the value  $g^{(2)}(0)$ for  the $DS$ state is
\be
g^{(2)}(0)=1+[\frac{2|\nu|^2}{\langle  n_0 \rangle }+\frac{|\nu|^2+|z_0|^2 \mu |\nu|\cos(\phi-\phi_z)}{{\langle  n_0 \rangle}^2}]\label{g2}
\ee
and the value  $g^{(3)}(0)$ for the $DS$ state is
\bq
g^{(3)}(0) &=& 1+[\frac{6|\nu|^2}{\langle  n_0 \rangle }+\frac{3|\nu|^2(4|\nu|^2+3)}{{\langle  n_0 \rangle}^2}\nonumber\\
&&{}-\frac{4|\nu|^6-3(|z_0|^2+3|\nu|^2)|z_0|^2 \mu |\nu|\cos(\phi-\phi_z)}{{\langle  n_0 \rangle}^3}]\label{g3}
\eq

These results [Eq.(\ref{g2}) and Eq.(\ref{g3})] are plotted in Figure 1 and Figure 2. From the figures we see that the experimental results are consistent with values of $r$ between 0 and 4 ($ r=0 $ is a pure condensate $D$ state).


In this note we have constructed a state for a Bose-Einstein condensate based on the $su(1,1)$ spectrum-generating algebra structure of the mean-field hamiltonian, and the Heisenberg-Weyl coherent state structure which gives non-vanishing boson operator expectations. It is a common feature of mean-field approximations that these give rise to the loss of conserved quantities (in our case loss of number conservation) and the consequent appearance  of associated order parameters, which here  are $<a_0>$ and $<a_k a_{-k}>$. This is a general property of the algebraic approach\cite{order}. The question of loss of number conservation is considered in some detail by Girardeau\cite{Gir}. However the linearisation procedure herein adopted retains the momentum-conservation properties of the original hamiltonian, as in superfluidity and superconductivity. The resulting $DS$ state is similar to a squeezed coherent state, familiar in Quantum Optics, and will undoubtedly give rise to interesting squeezing phenomena which will be explored later. 
After completion of this work, it has been brought to our attention that a similar $DS$ description of the condensate was also obtained by Navez\cite{navez}   from slightly different premises. In this note we showed that the $DS$ state provides better fits to the experimental results on the  correlation functions associated with the BEC state.
Although it might be argued that the additional freedom inherent in the extra parameters associated with the $DS$ state must give better fits to the experiments, it should be noted that the changes to the coherent state values (of unity) are in one direction only (positive) and are therefore only consistent with experimental values greater than one.

%
\vspace{2cm}
\epsfxsize=10cm
\hspace{2cm}
\epsfbox{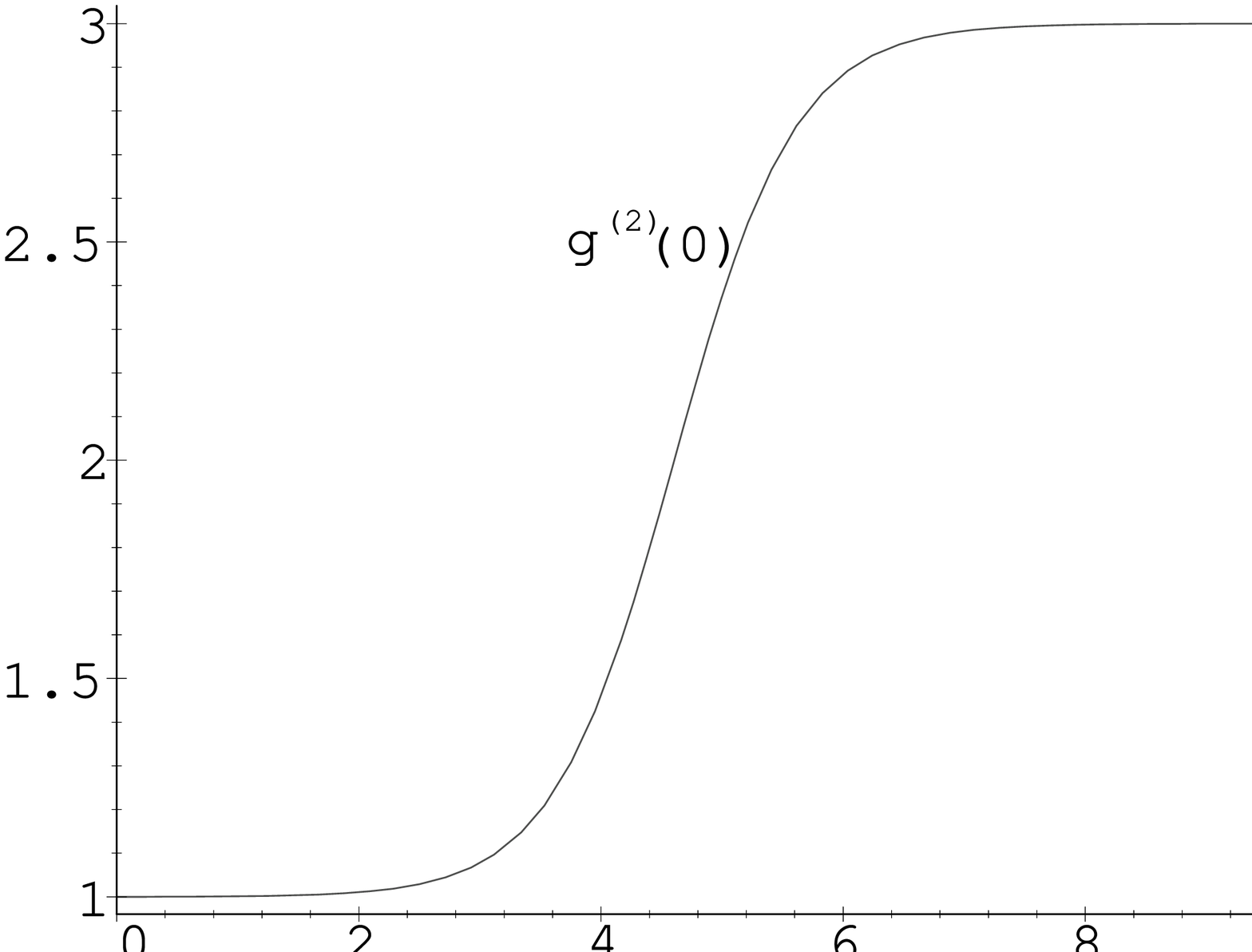}
\begin{figure}
\caption{ The Second-order Correlation Function $ {g^{(2)}(0)} $ for the $DS$ state  $\;$ $(|z_0|=50) $}
\end{figure}
\vspace{2cm}
\epsfxsize=10cm
\hspace{2cm}
\epsfbox{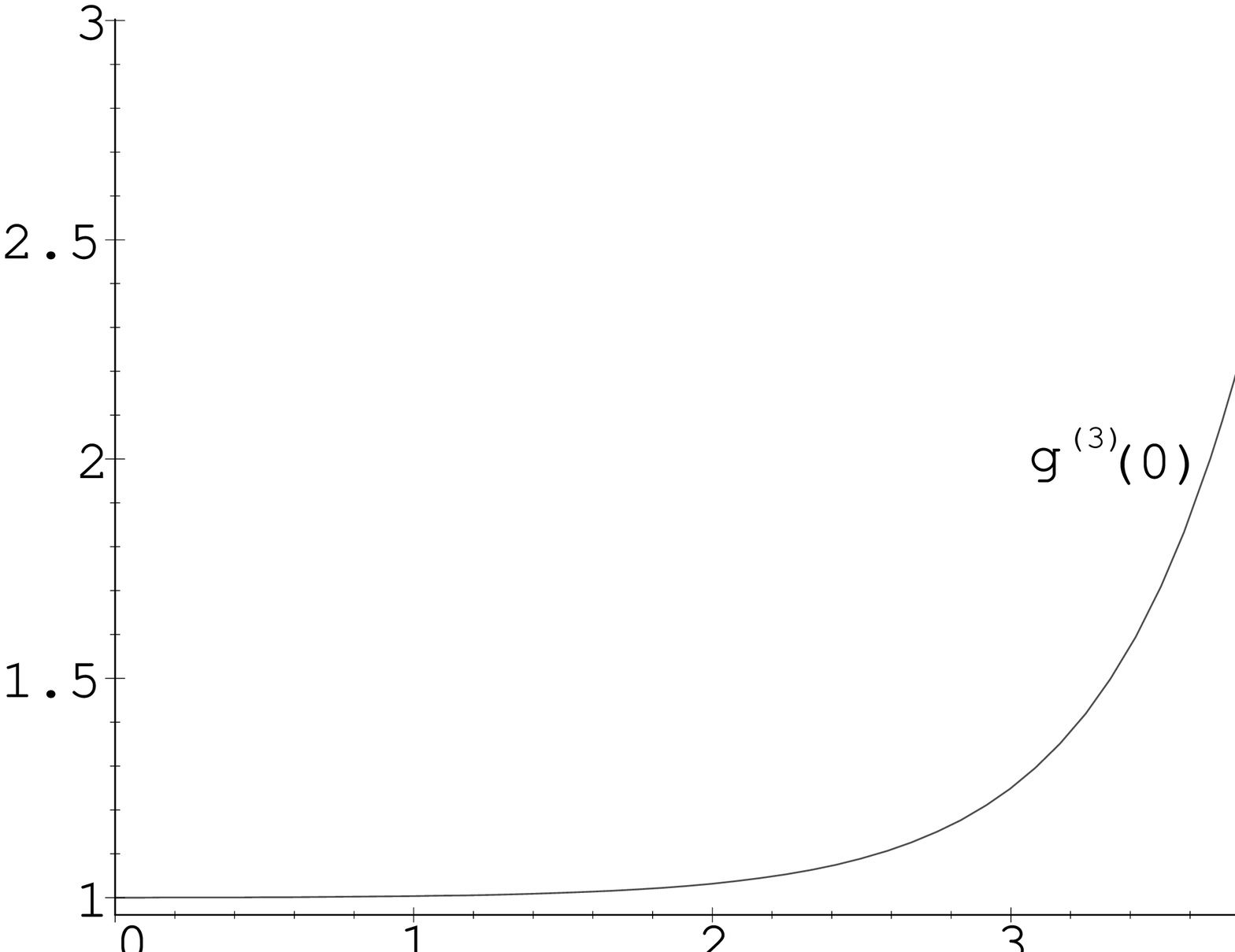}
\begin{figure}
\caption{ The Third-order Correlation Function $ {g^{(3)}(0)} $ for the $DS$ state  $\;$ $(|z_0|=50) $}
\end{figure}
\end{document}